\documentclass[11pt,twoside]{article}
\usepackage{asp2010}
\usepackage{amssymb}
\usepackage{graphicx,mathptm}
\newcommand{\be}{\begin{equation}}
\newcommand{\ee}{\end{equation}}
\newcommand{\bea}{\begin{eqnarray}}
\newcommand{\eea}{\end{eqnarray}}
\newcommand{\vc}[1]{\textbf{\em #1}}

\resetcounters

\begin{document}

\title{Astrophysical Reconnection and Particle Acceleration}
\author{ A. Lazarian$^1$, G. Kowal$^2$ \& B. Gouveia dal Pino$^2$
\affil{$^1$ Astronomy Department, University of Wisconsin, Madison, WI 53706, US}
\affil{$^2$ Instituto de Astronomia, Geof\'\i sica e Ci\^encias Atmosf\'ericas,
  Universidade de S\~ao Paulo, Rua do Mat\~ao, 1226 -- Cidade Universit\'{a}ria,
  CEP 05508-090, S\~ao Paulo/SP, Brazil}}

\begin{abstract}
Astrophysical reconnection takes place in a turbulent medium.  The turbulence in
most cases is pre-existing, not caused by the reconnection itself.  The model of
magnetic reconnection in Lazarian \& Vishniac (1999) predicts that in the
presence of turbulence the reconnection becomes fast, i.e. it is independent of
resistivity, but dependent on the level of turbulence.  Magnetic reconnection
injects energy into plasmas through a turbulent outflow from the reconnection
region and this outflow can enhance the level of turbulence creating bursts of
reconnection.  Magnetic reconnection in the presence of turbulence can
accelerate energetic particles through the first order Fermi mechanism, as was
predicted in Gouveia dal Pino \& Lazarian (2005).  We discuss new numerical
results on particle acceleration in turbulent reconnection, compare the
acceleration arising from turbulent reconnection to the acceleration of
energetic particles in turbulent medium.
\end{abstract}

\section{Reconnection and Numerical Studies}

It is generally believed that a magnetic field embedded in a highly conductive
fluid preserves its topology for all time due to the magnetic fields being
frozen-in \citep{alfven42,parker79}.  Although ionized astrophysical objects,
like stars and galactic disks, are almost perfectly conducting, they show
indications of changes in topology, ``magnetic reconnection'', on dynamical time
scales \cite[see][]{parker70}.  Reconnection can be observed directly in the
solar corona \cite[e.g.][]{yokoyama95}, but can also be inferred from the
existence of large-scale dynamo activity inside stellar interiors
\cite[see][]{parker93}.  Solar flares and $\gamma$-ray bursts
\cite[see][]{lazarian03,zhang11} are usually associated with magnetic
reconnection.  A lot of previous work has concentrated on showing how
reconnection can be rapid in plasmas with very small collisional rates
\cite{drake01,drake06}, which substantially constrains astrophysical
applications of the corresponding reconnection models.

A theory of magnetic reconnection is necessary to understand whether
reconnection is represented correctly in numerical simulations. One should keep
in mind that reconnection is fast in computer simulations due to high numerical
diffusivity. Therefore, if there are situations where magnetic fields reconnect
slowly, numerical simulations do not adequately reproduce the astrophysical
reality. This means that if collisionless reconnection is indeed the only way to
make reconnection fast, then the numerical simulations of many astrophysical
processes, including those in interstellar media, which is collisional at the
relevant scales, are in error. At the same time, it is not possible to conclude
that reconnection must always be fast on the empirical grounds, as solar flares
require periods of flux accumulation time, which correspond to slow
reconnection.

To understand the difference between reconnection in astrophysical situations
and in numerical simulations, one should recall that the dimensionless
combination that controls the resistive reconnection rate is the Lundquist
number\footnote{The magnetic Reynolds number, which is the ratio of the magnetic
field decay time to the eddy turnover time, is defined using the injection
velocity $v_l$ as a characteristic speed instead of the Alfv\'en speed $V_A$,
which is taken in the Lundquist number.}, defined as $S = L_xV_A / \lambda$,
where $L_x$ is the length of the reconnection layer, $V_A$ is the Alfv\'en
velocity, and $\lambda=\eta c^2/4\pi$ is Ohmic diffusivity. Because of the huge
astrophysical length-scales $L_x$ involved, the astrophysical Lundquist numbers
are also huge, e.g. for the ISM they are about $10^{16}$, while present-day MHD
simulations correspond to $S<10^4$. As the numerical efforts scale as $L_x^4$,
where $L_x$ is the size of the box, it is feasible neither at present nor in the
foreseeable future to have simulations with realistically Lundquist numbers.

\section{Reconnection and Turbulence}

While astrophysical fluids show a wide variety of properties in terms of their
collisionality, degree of ionization, temperature etc., they share a common
property, namely, most of the fluids are turbulent. The turbulent state of the
fluids arises from large Reynolds numbers $Re\equiv LV/\nu$, where $L$ is the
scale of the flow, $V$ is it velocity and $\nu$ is the viscosity, associated
with astrophysical media. Note, that the large magnitude of $Re$ is mostly the
consequence of the large astrophysical scales $L$ involved as well as the fact
that (the field-perpendicular) viscosity is constrained by the presence of
magnetic field.

Observations of the interstellar medium reveal a Kolmogorov spectrum of electron
density fluctuations \cite[see][]{armstrong95,chepurnov10} as well as steeper
spectral slopes of supersonic velocity fluctuations \cite[see][for
review]{lazarian09}. Measurement of the solar wind fluctuations also reveal
turbulence power spectrum \citep{leamon98}. Ubiquitous non-thermal broadening of
spectral lines as well as measures obtained by other techniques
\cite[see][]{burkhart10,gaensler11} confirm that turbulence is present
everywhere we test for its existence.  As turbulence is known to change many
processes, in particular the process of diffusion, the natural question is how
it affects magnetic reconnection.

To deal with strong, dynamically important magnetic fields \citet[][henceforth
LV99]{lazarian99} proposed a model of fast reconnection in the presence of
sub-Alfv\'enic turbulence. It is important to stress that unlike laboratory
controlled settings, in astrophysical situations turbulence is preexisting,
arising usually from the processes different from reconnection itself. In fact,
any modeling of astrophysical reconnection should account for the fact that
magnetic reconnection takes place in the turbulent environment and in most cases
the turbulence does not arise from magnetic reconnection. The analogy here can
be as follows: turbulence that is experienced by the plane does not arise from
the plane motion, but preexist in the atmosphere.

LV99 identified stochastic wandering of the magnetic field-lines as the most
critical property of MHD turbulence which permits fast reconnection and obtained
analytical relations between the reconnection rate and the turbulence intensity
and the turbulence injection scale.

\begin{figure}[ht]
 \center
 \includegraphics[width=0.4\columnwidth]{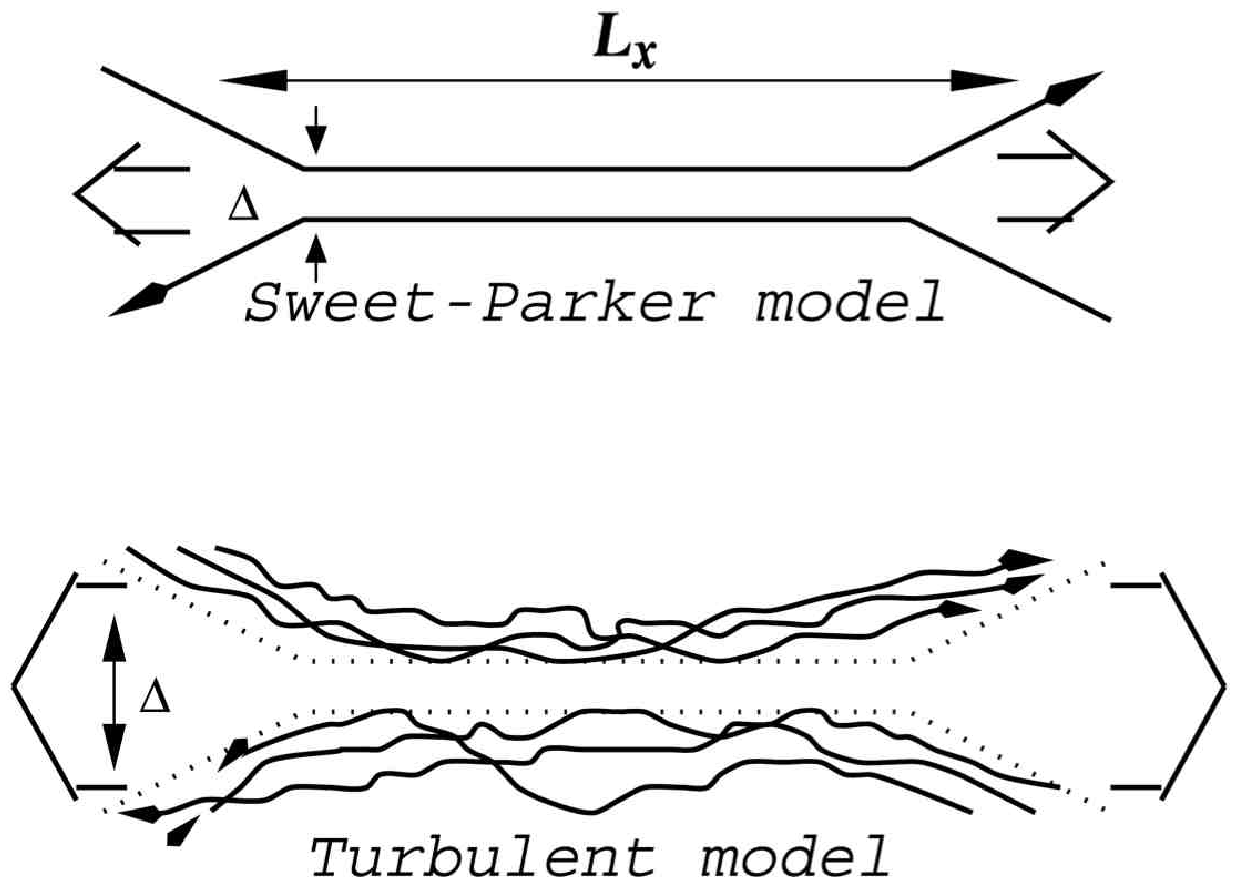}
 \includegraphics[width=0.4\columnwidth]{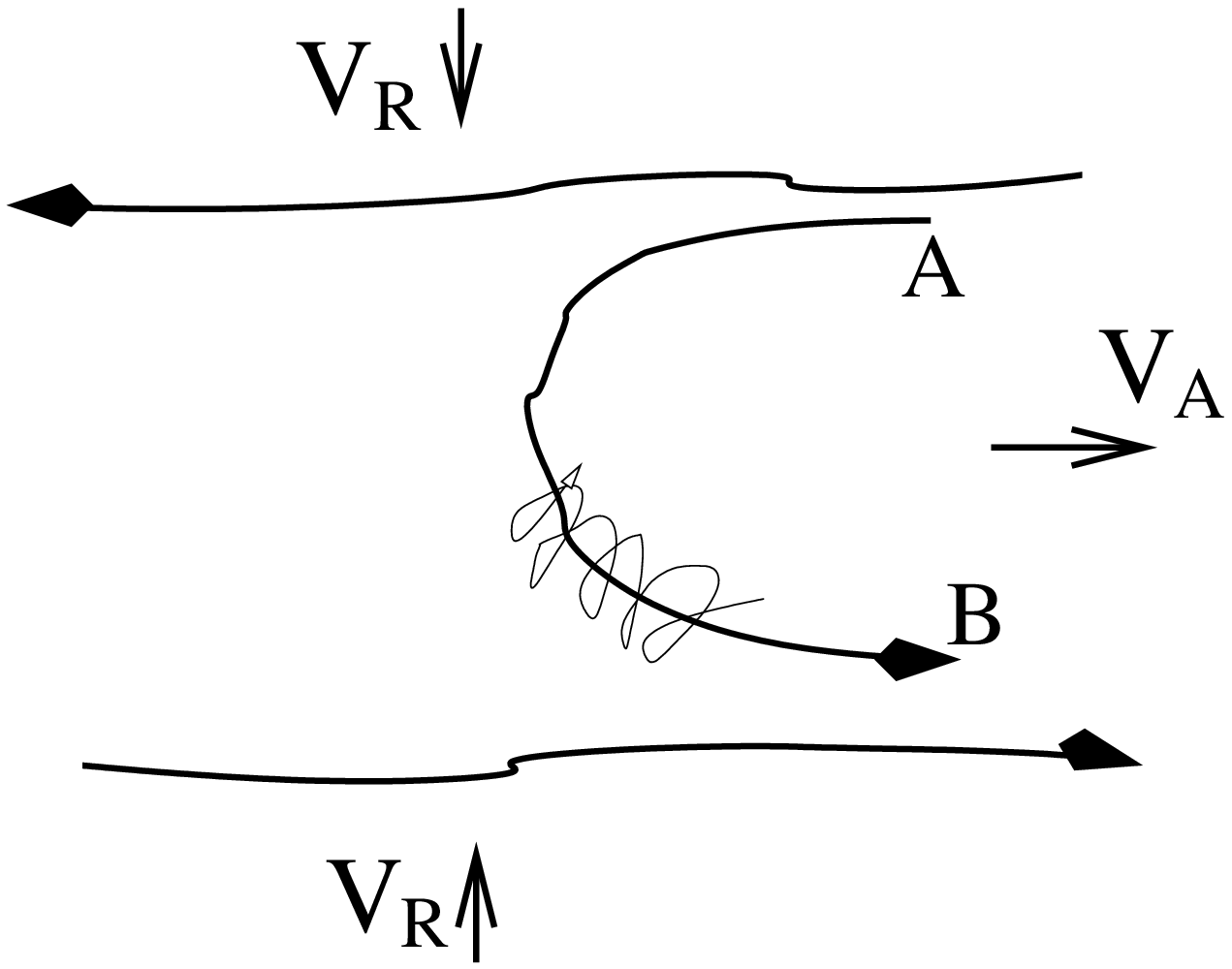}
 \caption{Left panel. Sweet-Parker model versus the model in LV99. Turbulence
makes the outflow region much wider and independent of resistivity. From
\cite{lazarian04}. Right panel. Acceleration of energetic particles as the
magnetic field shrinks as the result of reconnection. From Lazarian 2005.}
\end{figure}

It worth noting that the LV99 model (see Fig. 1, left panel) is radically
different from its predecessors which also appealed to the effects of
turbulence.  For instance, unlike \cite{speiser70} and \cite{jacobson84}Jacobson
(1984) the model does not appeal to changes of the microscopic properties of the
plasma. The nearest progenitor to LV99 was the work of
\cite{matthaeus85,matthaeus86}, who studied the problem numerically in 2D MHD
and who suggested that magnetic reconnection may be fast due to a number of
turbulence effects, e.g. multiple X points and turbulent EMF.  However,
\cite{matthaeus85,matthaeus86} did not realize the key role played by magnetic
field-line wandering, and did not obtain a quantitative prediction for the
reconnection rate, as did LV99.

LV99 revealed a very intimate relation between turbulence and magnetic
reconnection. First of all, it shows that reconnection is a necessary ingredient
of MHD turbulence, this is the process that makes the currently accepted picture
of MHD turbulence \cite{goldreich95} self-consistent.  Moreover, further
research in \citet[][henceforth ELV11]{eyink11} revealed that the expressions of
reconnection rate in LV99 can be obtained from the basic fluid turbulence
concept of Richardson diffusion.

\section{Reconnection and Plasma effects}

For years plasma effects have been considered essential for fast magnetic
reconnection. On the contrary, LV99 makes use of the MHD approximation. This
raises the issue to what extend the LV99 model is applicable to astrophysical
plasmas.

An MHD description of plasmas has been revisited recently in ELV11. There three
characteristic length-scales were considered: the ion gyroradius $\rho_i,$ the ion
mean-free-path length $\ell_{mfp,i}$, and the scale $L$ of large-scale variation
of magnetic and velocity fields. Astrophysical plasmas are in many cases
``strongly collisional'' in the sense that $\ell_{mfp,i}\ll \rho_i,$ which is the case for
the interiors of stars and accretion disks. In such cases, a fluid description of
the plasma is valid. In the ``weakly collisional''  $\ell_{mfp,i}\gg \rho_i,$.
The ratio
\begin{equation}
\frac{\ell_{mfp,i}}{\rho_i}\propto \frac{\Lambda}{\ln\Lambda}\frac{v_A}{c},
\end{equation}
follows from the standard formula for the Coulomb collision frequency (see
\cite{fitzpatrick11}, eq.(1.25)). Here $\Lambda=4\pi n\lambda_D^3$ is the plasma
parameter, or number of particles within the Debye screening sphere. When
astrophysical plasmas are very weakly coupled (hot and rarefied), then $\Lambda$
is large, e.g. of the order of $10^9$ or more for the warm component of the
interstellar medium or solar wind (see Table~1 in ELV11). For such ratio the
expansion over small ion Larmor radius $\rho_i$ provides ``kinetic MHD equations'' which
differ from the standard MHD by having anisotropic pressure tensor.

Plasmas that are not strongly collisional can be divided into two cases:
``collisionless'' plasmas for which $\ell_{mfp,i}\gg L,$ the largest scales of
interest, and ``weakly collisional'' plasmas for which $L\gg \ell_{mfp,i}.$ In
the latter case the``kinetic MHD'' description can be further reduced in
complexity at  scales greater than $\ell_{mfp,i}$ (see ELV11). This reproduces a
fully hydrodynamic MHD description at those scales.  For instance, the warm
ionized ISM is ``weakly collisional'', while post-CME current sheets and the
solar wind impinging on the magnetosphere are ``collisionless.''

Plasmas that are not strongly collisional further divide into two cases:
``collisionless'' plasmas for which $\ell_{mfp,i}\gg L,$ the largest scales of
interest, and ``weakly collisional'' plasmas for which $L\gg \ell_{mfp,i}.$ In
the latter case the``kinetic MHD'' description can be further reduced in
complexity at  scales greater than $\ell_{mfp,i}$ (see ELV11). This reproduces
a fully hydrodynamic MHD description at those scales, with anisotropic transport
behavior associated to the well-magnetized limit.  Among our examples in Table 1
above, the warm ionized ISM is ``weakly collisional'', while post-CME current
sheets and the solar wind impinging on the magnetosphere are close to being
``collisionless.''

Additional important simplifications occur if the following assumptions are
satisfied:  turbulent fluctuations are small compared to the mean magnetic
field,  have length-scales parallel to the mean field much larger than
perpendicular length-scales, and have frequencies low compared to the ion
cyclotron frequency.  These are standard assumptions of the \cite{goldreich95}
theory of MHD turbulence.  They are the basis of the ``gyrokinetic
approximation''  \citep{schekochihin07,schekochihin09}.  At length-scales
larger than the Larmor radius $\rho_i$, another reduction takes place. The
incompressible shear-Alfven wave modes exhibit dynamics independent of 
compressive motions and can be described by the ``Reduced MHD'' (RMHD)
equations \cite[see][]{goldreich95,cho03}.  This fact is essential 
for the LV99 justifying the use of the treatment based on an
incompressible MHD model.

Within the LV99 model, the reconnection rate is determined by large scale magnetoc wandering (see Figure 1), while small scale plasma effects may change the local reconnection which are irrelevant for the global reconnection at least in the fully ionized plasma \cite[see][]{lazarian04}.  This conclusion is supported by simulations in \cite{kowal09} where plasma effect were simulated by using anomalous resistivity. We should mention that although plasma effects do not change the global reconnection rate, can be important for the acceleration of electrons.

While Hall MHD is a default for many researchers, ELV11 showed that the effects of the Hall term on the field wandering
is negligible on scales larger than $\rho_i$ even if the Hall term in the generalized Ohm equation is dominant.

\section{Reconnection and Particle Acceleration}

Magnetic reconnection results in shrinking of magnetic loops and the charged particles entrained over magnetic loops get accelerated (see Figure 1, right panel).  This process was proposed in \citet[][henceforth GL05]{degouveia05} (see
also Lazarian 2005, 2006) for the LV99 reconnection and then was adopted for the collisionless reconnection in \cite{drake06}.  The physics of the acceleration is the same although GL05 appealed to the 3D magnetic bundles (see Figure 1), while \cite{drake06} considered 2D shrinking islands.  The latter is the consequence of the constrained 2D geometry and they present a strongly degenerate
case in 3D. The difference in dimensions affects the acceleration efficiency according to \cite{kowal11}.

GL05 claimed that the acceleration is of the first order Fermi type. This was tested recently in \cite{kowal12}.  Below we descibe the numerical set up and the results of calculations.

In order to integrate the test particle trajectories we freeze in time a data
cube obtained from the MHD models of reconnection and turbulence performed in
\cite{kowal09} and inject test particles in the domain with random initial
positions and directions and with an initial thermal distribution. For each
particle we solve the relativistic motion equation
\begin{equation} \frac{d}{d
t} \left( \gamma m \vc{u} \right) = q \left( \vc{E} + \vc{u} \times \vc{B} \right) , \label{eq:ptrajectory}
\end{equation}
where $m$, $q$ and $\vc{u}$ are the particle mass, electric charge and velocity,
respectively, $\vc{E}$ and $\vc{B}$ are the electric and magnetic fields,
respectively, $\gamma \equiv \left( 1 - u^2 / c^2 \right)^{-1}$ is the Lorentz
factor, and $c$ is the speed of light.  The electric field $\vc{E}$ is taken
from the MHD simulations $\vc{E} = - \vc{v} \times \vc{B} + \eta \vc{J}$, where
$\vc{v}$ is the plasma velocity, $\vc{J} \equiv \nabla \times \vc{B}$ is the
current density, and $\eta$ is the Ohmic resistivity coefficient.  We neglect
the resistive term above since its effect on particle acceleration is negligible
\cite{kowal11}.

\begin{figure}[ht]
 \center
 \includegraphics[width=0.48\textwidth]{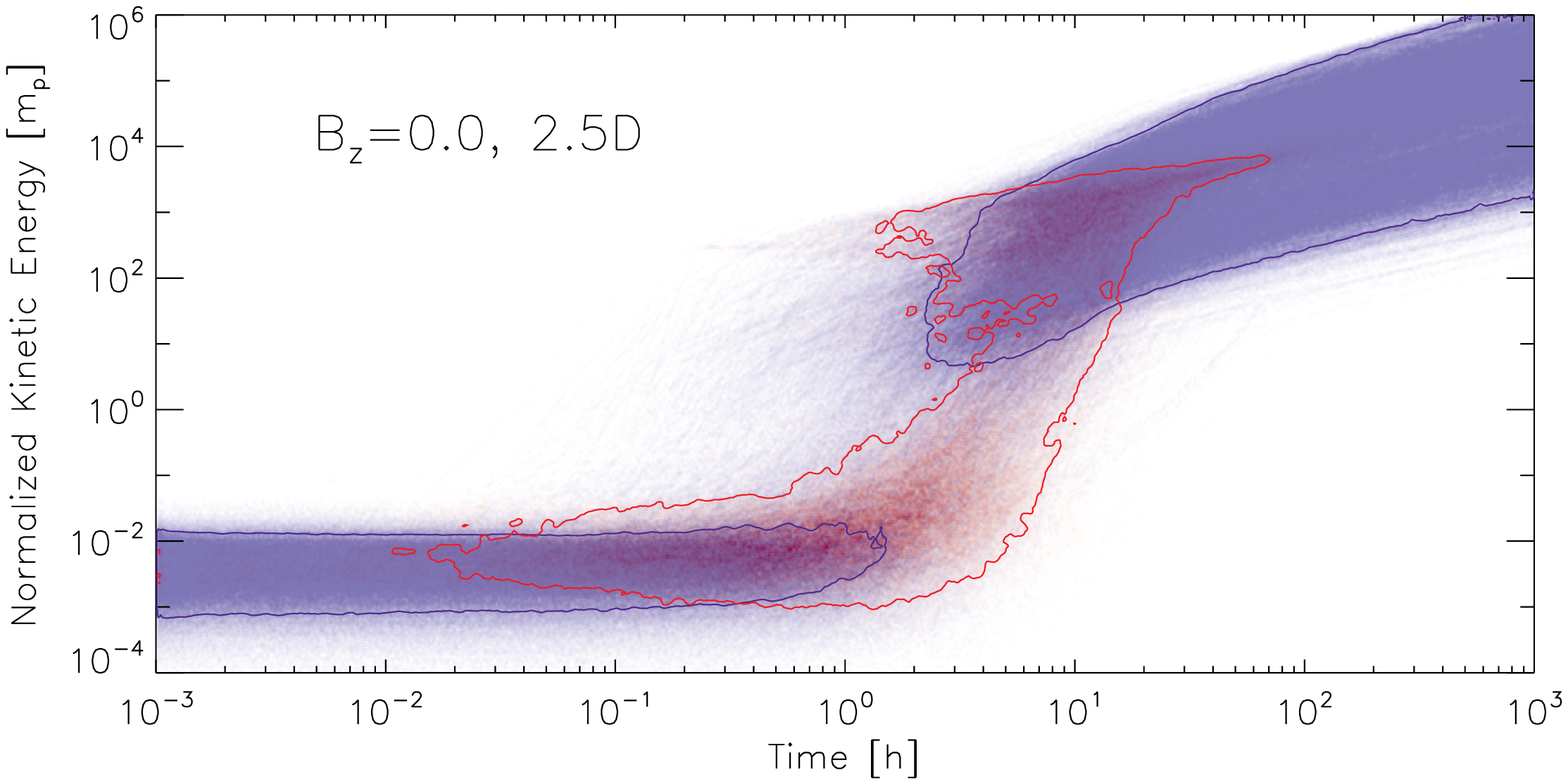}
 \includegraphics[width=0.48\textwidth]{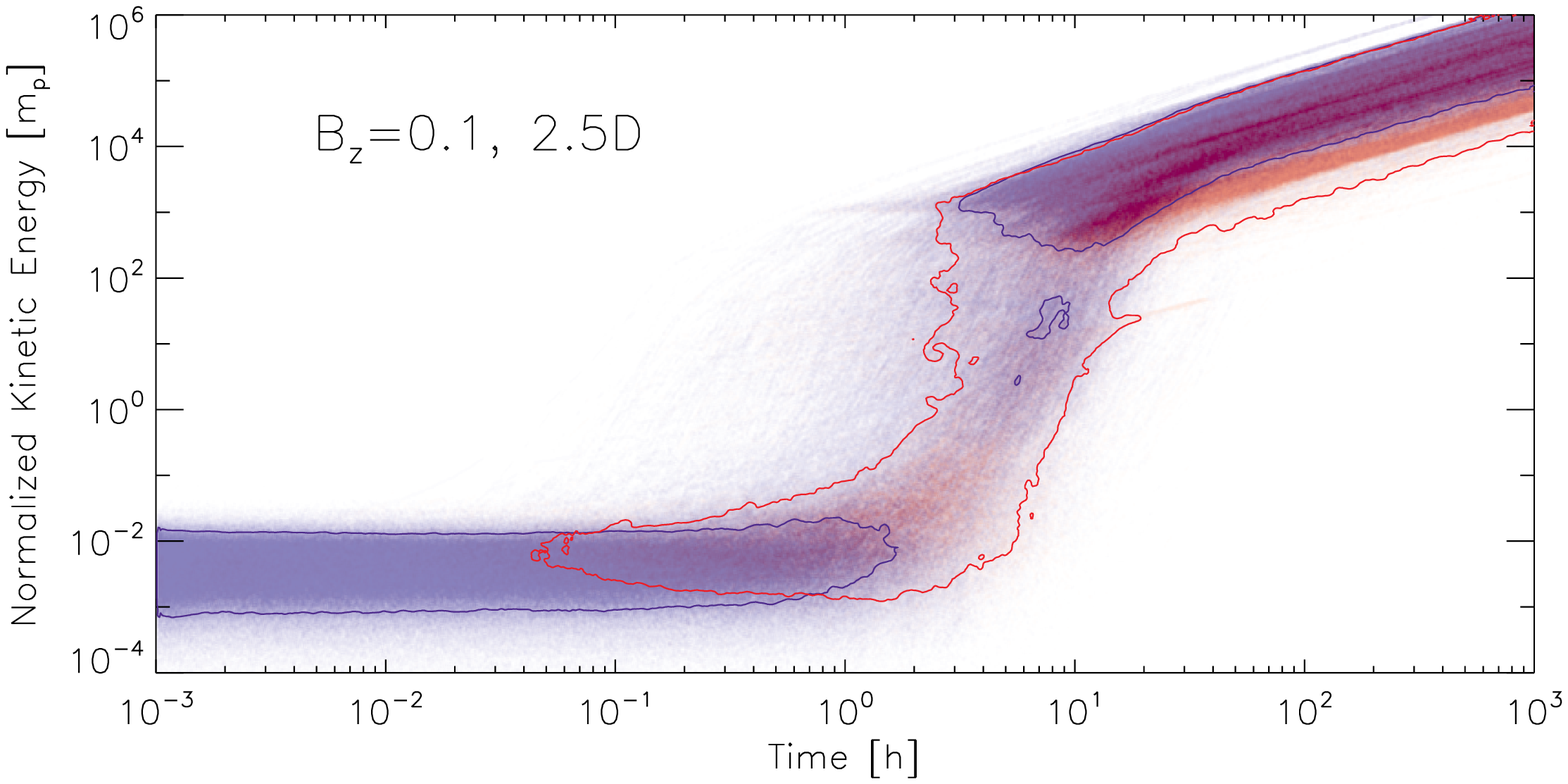}
 \includegraphics[width=0.48\textwidth]{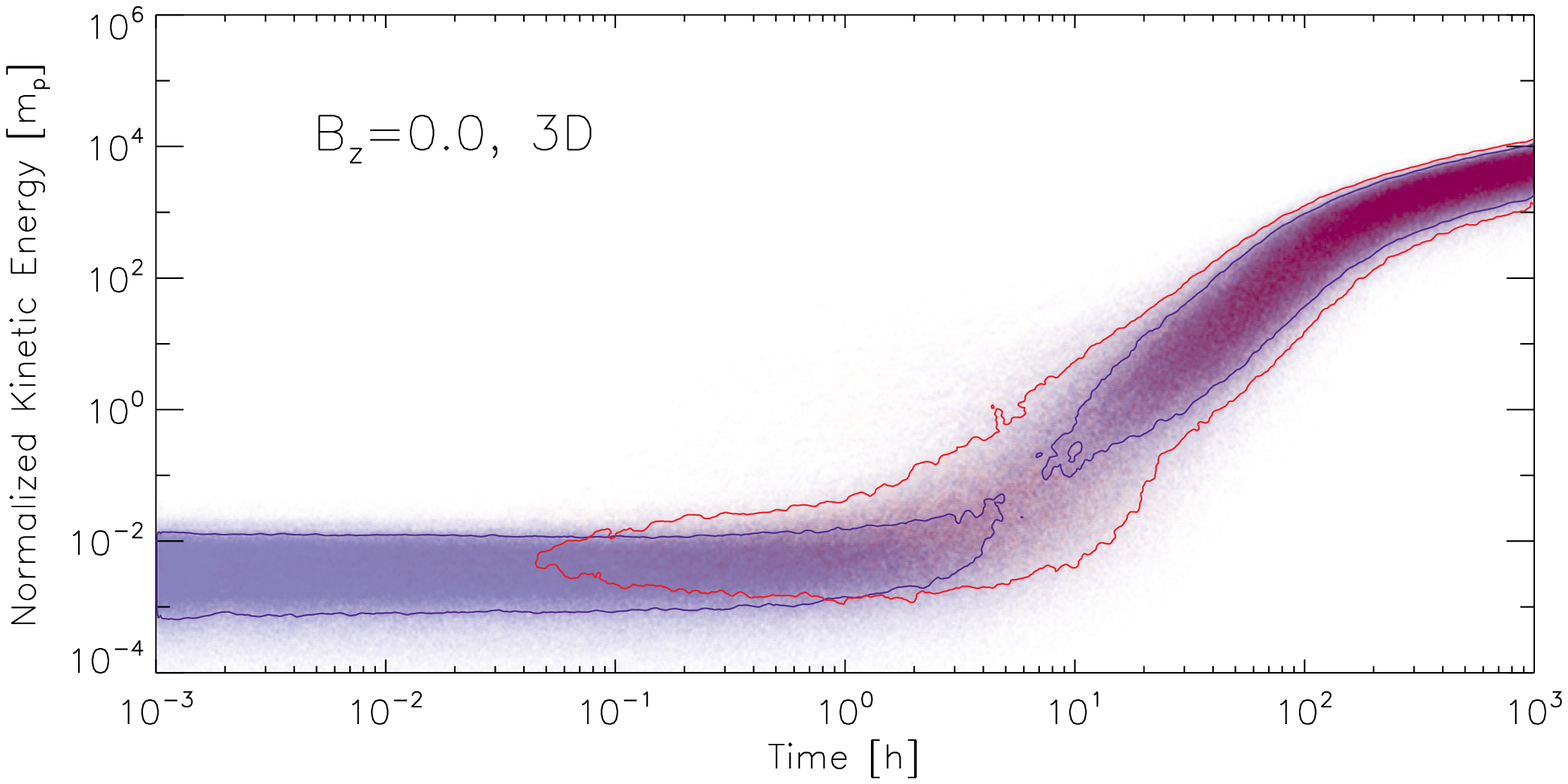}
 \caption{Kinetic energy evolution of a group of 10$^4$ protons in 2D models of
reconnection with a guide field $B_z$=0.0 and 0.1 (top panels, respectively).
In the bottom panel a fully 3D model with initial $B_z$=0.0 is presented.  The
colors show how the parallel (red) and perpendicular (blue) components of the
particle velocities increase with time. The contours correspond to values 0.1
and 0.6 of the maximum number of particles for the parallel and perpendicular
accelerations, respectively.  The energy is normalized by the rest proton mass
energy.  The background magnetized flow with multiple current sheet layers is at
time 4.0 in Alfv\'en time units in all models. From \cite{kowal11}
\label{fig:energy_2d_3d}}
\end{figure}

In Figure~\ref{fig:energy_2d_3d}, we present the time evolution of the kinetic
energy of the particles which have their parallel and perpendicular (red and
blue points, respectively) velocity components accelerated for three models of
reconnection.  The upper left panel shows the energy evolution for a 2D model
without the guide field (as in the models studied in the previous sections).
Initially, the particles pre-accelerate by increasing their perpendicular
velocity component only.  Later we observe an exponential growth of energy
mostly due to the acceleration of the parallel component which stops after the
energy reaches values of 10$^3$--10$^4$~$m_p$ (where $m_p$ is the proton rest
mass energy).  Further on, particles accelerate their perpendicular component
only with smaller linear rate in a log-log diagram.  In 2.5D case, there is also
an initial slow acceleration of the perpendicular component followed by the
exponential acceleration of the parallel velocity component.  Due to the
presence of a weak guide field, the parallel component accelerates further to
higher energies at a similar rate as the perpendicular one.  This implies that
the presence of a guide field removes the restriction seen in the 2D model
without a guide field and allows the particles to increase their parallel
velocity components as they travel along the guide field, in open loops rather
than in confined 2D islands.  This result is reassured by the 3D model in the
bottom panel of Figure~\ref{fig:energy_2d_3d}, where no guide field is necessary
as the MHD domain in fully three-dimensional.  In this case, we clearly see a
continuous increase of both components, which suggests that the particle
acceleration behavior changes significantly when 3D effects are considered,
where open loops replace the closed 2D reconnecting islands.

\section{Conclusions}

The results of these studies can be very briefly summarized as follows:

1. Advances in the understanding of magnetic reconnection in the MHD regime, in
particular, turbulent magnetic reconnection in \cite{lazarian09} model motivates
the studies of whether the reconnection in this regime can accelerate energetic
particles.

2. Contracting magnetic loops in magnetic reconnection in 2D, in the MHD regime,
provides the acceleration analogous to that observed in PIC simulations, which
proves that the acceleration in reconnection regions is a universal process
which is not determined by the details of plasma physics.

3. Acceleration of energetic particles in 2D and 3D shows substantial
differences, which call for focusing on realistic 3D geometries of reconnection.
Our study also shows that apart from the first order Fermi acceleration,
additional acceleration processes may occur within reconnection sites.

\paragraph{Acknowledgements.}

The research of AL is supported
by the Center for Magnetic Self-Organization in Laboratory and Astrophysical
Plasmas and NSF Grants AST-08-08118 and NASA grant A0000090101.  AL also
acknowledged Humboldt Award at the Universities of Cologne and Bochum, as well
as a Fellowship at the International Institute of Physics (Brazil). GK and EMGDP 
acknowledge the support by the FAPESP grants no. 2006/50654-3 and
2009/50053-8, and the CNPq  grant no. 300083/94-7.   This
research was also supported by the project TG-AST080005N through TeraGrid
resources provided by Texas Advanced Computing Center
(http://www.tacc.utexas.edu).


\begin{thebibliography}
\bibitem[Alfv{\'e}n(1942)]{alfven42}
 Alfv{\'e}n, H., {\it Existence of Electromagnetic-Hydrodynamic Waves}, \nat, 1942, 150, 405--406
\bibitem[Armstrong et al.(1995)]{armstrong95}
 Armstrong, J.~W., Rickett, B.~J., \& Spangler, S.~R., {\it Electron density power spectrum in the local interstellar medium}, 1995, \apj, 443, 209--221
\bibitem[Burkhart et al.(2010)]{burkhart10}
 Burkhart, B., Stanimirovi{\'c}, S., Lazarian, A., and Kowal, G., {\it Characterizing Magnetohydrodynamic Turbulence in the Small Magellanic Cloud}, 2010, \apj, 708, 1204--1220
\bibitem[Chepurnov and Lazarian(2010)]{chepurnov10}
 Chepurnov, A. \& Lazarian, A., {\it Extending the Big Power Law in the Sky with Turbulence Spectra from Wisconsin H{$\alpha$} Mapper Data}, 2010, \apj, 710,
853--858
\bibitem[Cho and Lazarian(2003)]{cho03}
 Cho, J.~and Lazarian, A., {\it Compressible magnetohydrodynamic turbulence: mode coupling, scaling relations, anisotropy, viscosity-damped regime and astrophysical implications}, 2003, \mnras, 345, 325--339
\bibitem[de Gouveia dal Pino \& Lazarian(2005)]{degouveia05}
 de Gouveia dal Pino, E.~M., \& Lazarian, A., {\it Production of the large scale superluminal ejections of the microquasar GRS 1915+105 by violent magnetic reconnection}, \aap, 2005, 441, 845--853
\bibitem[Drake(2001)]{drake01}
 Drake, J.~F., {\it Magnetic explosions in space}, 2001, \nat, 410, 525--526
\bibitem[Drake et al.(2006)]{drake06}
 Drake, J.~F., Swisdak, M., Schoeffler, K.~M., Rogers, B.~N., \& Kobayashi,
S., {\it Formation of secondary islands during magnetic reconnection}, 2006, GeoRL, 33,  13105.
\bibitem[Eyink et al.(2011)]{eyink11}
Eyink, G.~L., Lazarian, A., \& Vishniac, E.~T., {\it Fast Magnetic Reconnection and Spontaneous Stochasticity}, 2011, \apj, 743, 51
\bibitem{} Fitzpatrick, R. "Introduction to Plasma Physics, 
lecture notes
http://farside.ph.utexas.edu/teaching/plasma/plasm.html
\bibitem[Gaensler et al.(2011)]{gaensler11}
 Gaensler, B.~M., Haverkorn, M., Burkhart, B., Newton-McGee, K.~J., Ekers, R.~D., Lazarian, A., McClure-Griffiths, N.~M., Robishaw, T., Dickey, J.~M., \& Green, A.~J., {\it Low-Mach-number turbulence in interstellar gas revealed by radio polarization gradients}, 2011, \nat, 478, 214--217
\bibitem[Goldreich and Sridhar(1995)]{goldreich95}
 Goldreich, P. \& Sridhar, S., {\it Toward a theory of interstellar turbulence. 2: Strong alfvenic turbulence}, 1995, \apj, 438, 763--775
\bibitem[Jacobson(1984)]{jacobson84}
 Jacobson, A.~R., {\it A possible plasma-dynamo mechanism driven by particle transport}, 1984, Physics of Fluids, 27, 7--9
\bibitem[Kowal et al.(2009)]{kowal09}
 Kowal, G., Lazarian, A., Vishniac, E.~T., \& Otmianowska-Mazur, K., {\it Numerical Tests of Fast Reconnection in Weakly Stochastic Magnetic Fields}, 2009, \apj, 700, 63--85
\bibitem[Kowal et al.(2011)]{kowal11}
 Kowal, G., de Gouveia Dal Pino, E.~M., \& Lazarian, A., {\it Magnetohydrodynamic Simulations of Reconnection and Particle Acceleration: Three-dimensional Effects}, 2011, \apj, 735, 102
\bibitem[Kowal et al.(2012)]{kowal12}
 Kowal, G., de Gouveia Dal Pino, E.~M., \& Lazarian, A., {\it Acceleration in Turbulence and Weakly Stochastic Reconnection}, 2012, Physical Review Letters, in press
\bibitem{} \bibitem[Lazarian(2005)]{2005AIPC..784...42L} Lazarian, A.\ 2005, {\it Astrophysical Implications of Turbulent Reconnection: from cosmic rays to star formation}, in Magnetic 
Fields in the Universe: From Laboratory and Stars to Primordial 
Structures., 784, 42
\bibitem[Lazarian(2006)]{lazarian06}
 Lazarian, A., {\it Theoretical approaches to particle propagation and acceleration in turbulent intergalactic medium}, 2006, Astronomische Nachrichten, 327, 609
\bibitem[Lazarian(2009)]{lazarian09}
 Lazarian, A., {\it Obtaining Spectra of Turbulent Velocity from Observations}, 2009, \ssr, 143, 357--385
\bibitem[Lazarian \& Vishniac(1999)]{lazarian99}
 Lazarian \& Vishniac, {\it Reconnection in a Weakly Stochastic Field}, 1999, \apj, 517, 700--718
\bibitem[Lazarian et al.(2003)]{lazarian03}
 Lazarian, A., Petrosian, V., Yan, H., \& Cho, J., {\it Physics of Gamma-Ray Bursts: Turbulence, Energy Transfer and Reconnection}, 2003, arXiv:astro-ph/0301181
\bibitem[Lazarian et al.(2004)]{lazarian04}
 Lazarian, A., Vishniac, E.~T., \& Cho, J., {\it Magnetic Field Structure and Stochastic Reconnection in a Partially Ionized Gas}, 2004, \apj, 603, 180--197
\bibitem[Leamon et al.(1998)]{leamon98}
 Leamon, R.~J., Smith, C.~W., Ness, N.~F., Matthaeus, W.~H., \& Wong, H.~K., {\it Observational constraints on the dynamics of the interplanetary magnetic field dissipation range}, 1998, JGR, 103, 4775
\bibitem[Matthaeus and Lamkin(1985)]{matthaeus85}
 Matthaeus, W.~H. \& Lamkin, S.~L., {\it Rapid magnetic reconnection caused by finite amplitude fluctuations}, 1985, Physics of Fluids, 28, 303--307
\bibitem[Matthaeus and Lamkin(1986)]{matthaeus86}
 Matthaeus, W.~H. \& Lamkin, S.~L., {\it Turbulent magnetic reconnection}, Physics of Fluids, 1986, 29, 2513--2534
\bibitem[Parker(1970)]{parker70}
 Parker, E.~N., {\it The Generation of Magnetic Fields in Astrophysical Bodies. I. The Dynamo Equations}, \apj, 1970, 162, 665--673.
\bibitem[Parker(1979)]{parker79}
 Parker, E.~N., {\it Cosmical magnetic fields: Their origin and their activity}, 1979, Oxford Clarendon Press; New York, Oxford University Press
\bibitem[Parker(1993)]{parker93}
 Parker, E.~N., {\it A solar dynamo surface wave at the interface between convection and nonuniform rotation}, 1993, \apj, 408, 707--719
\bibitem[Schekochihin et al.(2007)]{schekochihin07}
 Schekochihin, A.~A., Iskakov, A.~B., Cowley, S.~C., McWilliams, J.~C., Proctor, M.~R.~E., \& Yousef, T.~A., {\it Fluctuation dynamo and turbulent induction at low magnetic Prandtl numbers}, 2007, New Journal of Physics, 9, 300
\bibitem[Schekochihin et al.(2009)]{schekochihin09}
 Schekochihin, A.~A., Cowley, S.~C., Dorland, W., Hammett, G.~W., Howes, G.~G., Quataert, E., \& Tatsuno, T., {\it Astrophysical Gyrokinetics: Kinetic and Fluid Turbulent Cascades in Magnetized Weakly Collisional Plasmas}, 2009, \apjs, 182, 310--377
\bibitem[Speiser(1970)]{speiser70}
 Speiser, T.~W., {\it Conductivity without collisions or noise}, 1970, Planetary and Space Science, 18, 613
\bibitem[Yokoyama and Shibata(1995)]{yokoyama95}
 Yokoyama, T. \& Shibata, K., {\it Magnetic reconnection as the origin of X-ray jets and H{$\alpha$} surges on the Sun}, 1995, \nat, 375, 42--44
\bibitem[Zhang and Yan(2011)]{zhang11}
 Zhang, B.~and Yan, H., {\it The Internal-collision-induced Magnetic Reconnection and Turbulence (ICMART) Model of Gamma-ray Bursts}, 2011, \apj, 726, 90
\end{thebibliography}
\end{document}